\documentclass[12pt,numreferences]{kluwermodified}

\usepackage{dsfont}
\usepackage{epsfig}
\begin{document}

\begin{opening}
\title{Causal construction of the massless vertex diagram}
\author{ANDREAS \surname{ASTE}}
\institute{Department of Physics and Astronomy, Theory Division,
University of Basel, Klingelbergstrasse 82,\\
4056 Basel, Switzerland\\
e-mail: andreas.aste@unibas.ch}
\runningtitle{THE MASSLESS VERTEX DIAGRAM}
\runningauthor{ANDREAS ASTE}

\begin{abstract}
The massless one-loop vertex diagram is constructed by exploiting
the causal structure of the diagram in configuration space, which
can be translated directly into dispersive relations in momentum space.
\end{abstract}
\classification{AMS classification (2000)}{81T05,81T15,81T18.}

\end{opening}

\section{Introduction}
In the traditional approach to quantum field theory, one starts
from classical fields and a Lagrangean which describes the interaction.
These objects get quantized and $S$-matrix elements or Greens functions are
constructed with the help of the Feynman rules. 
A typical example which is used in this paper as a model theory
is the massless $\Phi^3$-theory, where the interaction
Hamiltonian density is given by the normal ordered third order monomial
of a free uncharged massless scalar field and a coupling
constant $\lambda$
\begin{equation}
{\cal{H}}_{int}(x)=\frac{\lambda}{3!} : \Phi(x)^3:,
\end{equation}
and $x$ is an element of Minkowski space, denoted by $\mathds{R}^4$ in the
following. The perturbative $S$-Matrix is then constructed according to the expansion
\begin{equation}
S={\bf 1}+\sum \limits_{n=1}^{\infty} \frac{(-i)^n}{n !} \int dx^4_1 ... dx^4_n
T\{{\cal{H}}_{int}(x_1) {\cal{H}}_{int}(x_2) \cdot ... \cdot {\cal{H}}_{int}(x_n)\}, \label{stoer1}
\end{equation}
where $T$ is the time-ordering operator.
It must be pointed out that the perturbation series eq. (\ref{stoer1}) is formal
and it is difficult to make any statement about the convergence of this series.
Furthermore, two problems arise in the expansion given above.
First, the time-ordered products
\begin{equation}
T_n(x_1,x_2, ...,x_n)=(-i)^n T\{{\cal{H}}_{int}(x_1) {\cal{H}}_{int}(x_2)
\cdot ... \cdot {\cal{H}}_{int}(x_n)\} \label{toprod}
\end{equation}
are usually plagued by ultraviolet divergences. However, these divergences can be removed by
regularization, such that the operator-valued distributions $T_n$ can be viewed as
well-defined, already regularized expressions.
Second, infrared divergences are also present in eq. (\ref{stoer1}).
This is not astonishing, since the $T_n$'s are operator-valued distributions, and
therefore must be smeared out by test functions in $\mathcal{S}(\mathds{R}^{4n})$,
the Schwartz space of functions of rapid decrease. One may therefore introduce
a test function $g(x) \in \mathcal{S}(\mathds{R}^{4})$ which plays the role of an
'adiabatic switching' and provides a cutoff in the long-range part of the interaction, which
can be considered as a natural infrared regulator \cite{GScharf,eg}.
Then the infrared regularized $S$-matrix is given by
\begin{equation}
S(g)={\bf 1}+\sum \limits_{n=1}^{\infty} \frac{1}{n !} \int dx^4_1 ... dx^4_n
T_n(x_1,...,x_n) g(x_1) \cdot ... \cdot g(x_n), \label{smoothedS}
\end{equation}
and an appropriate adiabatic limit $g \rightarrow 1$ must be performed at the end
of actual calculations in the right quantities (like cross sections)
 where this limit exists. This is not one of the standard strategies usually
found in the literature, however, it is the most natural one
in view of the mathematical framework used in perturbative quantum field
theory.

The vertex diagram appears in $\Phi^3$-theory at third order of perturbation theory
via
\begin{displaymath}
T_3(x_1,x_2,x_3)=\Biggl( \frac{-i \lambda}{3!} \Biggr)^3 T \{ :\Phi^3(x_1)::\Phi^3(x_2):
:\Phi^3(x_3): \}
\end{displaymath}
\begin{equation}
=\lambda^3 V(x_1,x_2,x_3) :\Phi(x_1)\Phi(x_2)\Phi(x_3): + 
{\mbox{\emph{other graphs}}},
\end{equation}
where $V(x_1,x_2,x_3)=\Delta_F(x_1-x_3) \Delta_F(x_3-x_2) \Delta_F(x_2-x_1)$ is given
by the scalar Feynman propagator discussed below in further detail.
Of course, the massless vertex functions plays also a prominent role in applied
quantum field theories like QCD, where it emerges in the three-gluon vertex \cite{Tarasov},
or in perturbative models of quantum gravity.

In this paper, the Fourier transform of the massless vertex distribution (also
called vertex function in the sequel)
\begin{displaymath}
\hat{V}(p,q)=\int d^4 x_1 d^4 x_2 V(x_1,x_2,x_3) e^{ip(x_1-x_3)+iq(x_3-x_2)}
\end{displaymath}
\begin{equation}
=(2 \pi)^{-4} \int \frac{d^4 k}{[(p-k)^2+i0][(q-k)^2+i0][k^2+i0]} \label{vertextrafo}
\end{equation}
will be calculated in a non-standard manner by making explicit use of causality
as a fundamental property of local quantum field theories \cite{PCT}. The presented dispersive
method is based on considerations in configuration space and integrals in
momentum space are not evaluated by the help of Wick rotations. Furthermore, no
mass terms are needed as regulators of infrared divergences, the vertex function is constructed
directly without any limiting procedure.

It should be noted that the vertex function is translation invariant in the sense that
$V(x_1,x_2,x_3)=V(x_1-a,x_2-a,x_3-a)$, where $a$ is an arbitrary four-vector.
This symmetry property can be inferred directly from the decomposition of the vertex function
into three Feynman propagators given above.
Introducing new integration variables $z_1=x_1-x_3$ and $z_2=x_3-x_2$, the Fourier transform
in eq. (\ref{vertextrafo}) can also be written as ($\Delta_F(x)=\Delta_F(-x)$)
\begin{displaymath}
\hat{V}(p,q)=\int d^4 z_1 d^4 z_2 \, V(z_1+x_3,-z_2+x_3,x_3) e^{ip z_1+iq z_2}
\end{displaymath}
\begin{equation}
=\int d^4 z_1 d^4 z_2  \, \Delta_F(z_1) \Delta_F(z_2) \Delta_F(z_1+z_2)
e^{ip z_1+iq z_2},
\end{equation}
which shows that $\hat{V}(p,q)$ does indeed not depend on $x_3$.

The focus of this work is on the causal and analytic properties of the vertex function
and its infrared-finite dispersive calculation in the massless case.
The more involved general-mass case of the vertex function has
been treated by 't Hooft and Veltman \cite{HooftVelt}, such that results for the
massless case could be obtained by considering the appropriate limit.
An explicit result for the massless case has been given in \cite{Ball},
and a discussion of the vertex function within a special framework based on a geometrical
interpretation of kinematic invariants and other quantities of one-loop diagrams is contained
in \cite{DavyGeom}, where a short overview over different calculational methods
for one-loop N-point functions can also be found.

Before investigating the causal structure of the scalar vertex function,
some important analytic properties of the Feynman propagator and other related
tempered distributions will be recalled in the following section.

\section{Properties of tempered distributions}
The Feynman propagator of a free scalar and charge neutral quantum field
$\Phi(x)$ fulfilling the wave equation
\begin{equation}
\Box \Phi(x) = \partial_\mu \partial^\mu \Phi(x)=0
\end{equation}
is given in configuration space by the vacuum expectation value
\begin{displaymath}
\Delta_F(x)=\Delta_F(x_1-x_2)=-i \langle 0 | T (\Phi(x) \Phi(0)) | 0 \rangle=
-i \langle 0 | T (\Phi(x_1) \Phi(x_2)) | 0 \rangle
\end{displaymath}
\begin{equation}
=\int \frac{d^4 k}{(2 \pi)^4} \frac{e^{-ikx}}{k^2+i0}
=\frac{i}{4 \pi^2} \frac{1}{x^2-i0} =
\frac{i}{4 \pi^2} P \frac{1}{x^2}-
\frac{1}{4 \pi} \delta (x^2),
\end{equation}
where $P$ denotes principal value
regularization and $\delta$ is the one-dimensional Dirac distribution depending
on $x^2=x_\mu x^\mu=(x^0)^2-(x^1)^2-(x^2)^2-(x^3)^2=x_0^2-{\vec{x}}^2$ \cite{Constantinescu}.

Knowledge of the basic properties of the Feynman propagator
is sufficient for many applications in perturbative quantum field theory.
For the forthcoming discussion, it is instructive
to review some of the useful properties
of basic tempered distributions appearing in quantum field theory.
A scalar neutral field $\Phi$ can be decomposed into a negative and positive frequency
part according to
\begin{equation}
\Phi(x)=\Phi^-(x)+\Phi^+(x)=
\int \frac{d^3k}{\sqrt{(2 \pi)^3 2 |\vec{k}|}} \Bigl[ a(\vec{k})
e^{-ikx}+a^\dagger(\vec{k}) e^{ikx} \Bigr] 
\end{equation}
with
\begin{equation}
[a(\vec{k}),a^\dagger(\vec{k}')]=\delta^{(3)}(\vec{k}-\vec{k}'), \quad
[a(\vec{k}),a(\vec{k}')]=[a^\dagger(\vec{k}),a^\dagger(\vec{k}')]=0.
\end{equation}
The commutation relations for such a field are given by the
positive and negative frequency Jordan-Pauli distributions
\begin{equation}
\Delta^\pm(x)=-i [\Phi^{\mp}(x) , \, \Phi^{\pm}(0)] =
-i \langle 0 |  [\Phi^{\mp}(x) , \, \Phi^{\pm}(0)] | 0 \rangle \, ,
\label{commutator}
\end{equation}
which have the Fourier transforms
\begin{equation}
\hat{\Delta}^{\pm}(k)= \int d^4x \, \Delta^{\pm}(x) e^{ikx}=
\mp (2 \pi i) \,  \Theta(\pm k^0) \delta(k^2) . \label{dfourier}
\end{equation}
The fact that the commutator
\begin{equation}
[\Phi(x),\Phi(0)]=i\Delta^+(x)+i\Delta^-(x) =: i \Delta(x)
\end{equation}
with the Fourier transform
\begin{equation}
\hat{\Delta}(k)=-(2 \pi i) \mbox{sgn}(k^0) \delta(k^2) \label{ftjp}
\end{equation}
vanishes for spacelike arguments $x^2 < 0$
due to the requirement of microcausality, leads to
the important property that the Jordan-Pauli distribution
$\Delta$ has {\emph{causal support}}, i.e. it vanishes outside the {\emph{closed}}
forward and backward light cone such that
\begin{equation}
\mbox{supp} \, \Delta(x) \subseteq \overline{V}^- \cup \overline{V}^+  \, , \quad
\overline{V}^\pm=\{x \, | \, x^2 \ge 0, \, \pm x^0 \ge 0 \}
\end{equation}
in the sense of distributions.
A further crucial observation is the fact that one can introduce the
retarded propagator $\Delta^{ret}(x)$ which coincides with $\Delta(x)$ on
$\overline{V}^+ \! - \{0 \}$, i.e. $\langle \Delta^{ret}, \varphi \rangle =
\langle \Delta, \varphi \rangle$ holds for all test functions in the
Schwartz space $\varphi \! \in
\! \mathcal{S}(\mathds{R}^4)$ with support $\mbox{supp} \, \varphi \subset
\mathds{R}^4 - \overline{V}^-$.
One may write in configuration space
\begin{equation}
\Delta^{ret}(x)=\Theta(x^0) \Delta(x) , \label{welldef}
\end{equation}
and transform this 'splitting' formula into a convolution in momentum space
\footnote{The Heaviside distribution $\Theta(x^0)$ could be replaced by
$\Theta(vx)$ with an arbitrary
vector $v \in V^+=\overline{V}^+ \! - \partial \overline{V}^+$
inside the open forward light cone.}
\begin{equation}
\hat{\Delta}^{ret}(k)= \int \frac{d^4 p}{(2 \pi)^4} \, \hat{\Delta}(p) \hat{\Theta}
(k-p). \label{convo}
\end{equation}
The Fourier transform of the Heaviside distribution $\Theta(x^0)$
can be calculated easily
\begin{equation}
\hat{\Theta}(k)=\lim_{\epsilon \rightarrow 0}
\int d^4x \, \Theta(x^0) e^{-\epsilon x^0} e^{ik_0x^0-i \vec{k}\vec{x}}=
\frac{(2 \pi)^3 i}{k^0+i0} \delta^{(3)}(\vec{k}) .
\end{equation} 
For the special case where $k$ is in the forward light cone $V^+$,
one can go to a Lorentz frame where $k=(k^0,\vec{0})$
such that eq.
(\ref{convo}) becomes
\begin{equation}
\hat{\Delta}^{ret}(k^0,\vec{0})=\frac{i}{2 \pi} \int dp^0 \frac{\hat{\Delta}
((p^0, {\vec{0}}))}{k^0-p^0+i0}=
\frac{i}{2 \pi} \int dt \, \frac{\hat{\Delta}
((tk^0, {\vec{0}}))}{1-t+i0} .
\end{equation}
Hence, for arbitrary $k \in V^+$, $\hat{\Delta}^{ret}$ would be
given by the {\emph{dispersion relation}}
\begin{equation}
\hat{\Delta}^{ret}(k)=\frac{i}{2 \pi} \int dt \, \frac{\hat{\Delta}
(tk)}{1-t+i0} . \label{disprel}
\end{equation}
However, the integral in eq. (\ref{disprel}) is undefined in the massless case
considered here. This illustrates the fact that it is not possible in general to
convert the product of two distributions into a convolution via a Fourier transform,
even though expression (\ref{welldef}) describes a well-defined distribution.
Here, one can circumvent this
problem, e.g., by introducing a mass for the field $\Phi$ as a regulator for the
corresponding massive Jordan-Pauli distribution which is also defined as a field
commutator (for a different method, see, e.g.,
\cite{Massless}). For the massive Jordan-Pauli distribution $\Delta_m(x)$ one gets
in momentum space
\begin{equation}
\hat{\Delta}_m(k)=-(2 \pi i) \mbox{sgn}(k^0) \delta(k^2-m^2) \label{ftmjp}
\end{equation}
and the following expression for the massive retarded distribution $\hat{\Delta}_m^{ret}(k)$
($k \! \in \! V^+$)
\begin{displaymath}
\hat{\Delta}_m^{ret}(k)=\int dt \, \frac{
\mbox{sgn}(tk^0) \delta(t^2k^2-m^2)}{1-t+i0}
\end{displaymath}
\begin{equation}
=\int dt \, \frac{
\bigl[ \delta(t-\frac{m}{\sqrt{k^2}})- \delta(t+\frac{m}{\sqrt{k^2}})
\bigr]}{2 \sqrt{k^2} m (1-t+i0) }=
\frac{1}{k^2-m^2} \quad (k^2 \ne m^2) \, .
\end{equation}
The $\delta$-distribution in the Fourier transformed Jordan-Pauli distribution
eq. (\ref{ftmjp}) contains a mass term which is not present in eq. (\ref{ftjp}),
such that $\hat{\Delta}_m(k)$ does not exhibit a singular behavior on the light cone
in momentum space.

As a special case of the edge of the wedge theorem \cite{PCT}
it is known that the Fourier transform of the retarded distribution
${\Delta}^{ret}(x)$ is the boundary value of an analytic function
$r(z)$, regular in $T^+ := \mathds{R}^4+ i V^+$ \cite{GScharf,Reeds}. This way one obtains
from $r(z)=1/(z^2-m^2)$, $z \! \in \! T^+$, and $m \rightarrow 0$
\begin{equation}
\hat{\Delta}^{ret}(k)=\frac{1}{k^2+ik^0 0} = \frac{1}{k^2+i0}+2 \pi i
\Theta(-k^0) \delta(k^2) . \label{composition}
\end{equation}
The analytic expression for the Feynman propagator is then recovered,
and one obtains
\begin{equation}
\hat{\Delta}^{ret}(k)=\hat{\Delta}_F(k)+\hat{\Delta}^-(k). \label{causal1}
\end{equation}

It is helpful for the forthcoming to understand eq. (\ref{causal1}). This is best done in
configuration space. One obviously has
\begin{displaymath}
i\Delta_F(x_1-x_2)=\langle 0 | T (\Phi(x_1) \Phi(x_2)) | 0 \rangle
\end{displaymath}
\begin{equation}
=\langle 0 | \Theta(x^0_1-x^0_2) \Phi(x_1) \Phi(x_2) | 0 \rangle+
\langle 0 | \Theta(x^0_2-x^0_1) \Phi(x_2) \Phi(x_1) | 0 \rangle
\end{equation}
and
\begin{displaymath}
i\Delta^-(x_1-x_2)=-\langle 0 | \Phi(x_2) \Phi(x_1) | 0 \rangle
\end{displaymath}
\begin{equation}
=-\langle 0 | \Theta(x^0_2-x^0_1) \Phi(x_2) \Phi(x_1) | 0 \rangle
-\langle 0 | \Theta(x^0_1-x^0_2) \Phi(x_2) \Phi(x_1) | 0 \rangle ,
\end{equation}
and therefore eq. (\ref{composition}) is fulfilled
\begin{displaymath}
i\Delta^{ret} (x_1-x_2)=i \Theta(x^0_1-x^0_2) \Delta(x_1-x_2)
\end{displaymath}
\begin{equation}
=\langle 0 | \Theta(x^0_1-x^0_2) \Phi(x_1) \Phi(x_2) | 0 \rangle
-\langle 0 | \Theta(x^0_1-x^0_2) \Phi(x_2) \Phi(x_1) | 0 \rangle .
\end{equation}
The discussion presented here can be applied the same way to second order
loop diagrams \cite{Aste,AsteSunrise,ASchw}. The problem of
ultraviolet divergences can be solved in such cases by using subtracted
dispersion relations \cite{GScharf}.

\section{Causality}
The causal properties of the Jordan-Pauli distribution and the retarded
propagator can be generalized in an analogous manner to the case
of the vertex function. For this purpose, one defines the
causal operator-valued three-point functions $\Delta_{3,op}^\pm$, $\Delta_3^{op}$
and related C-number valued contractions $\Delta_{3}^\pm$ and $\Delta_3$,
which are related to the vertex diagram via
\begin{displaymath}
\Delta_{3,op}^{-}(x_1,x_2,x_3)=
-T_2(x_2,x_3) T_1(x_1) - T_1(x_3) T_2(x_1, x_2) - T_2(x_1,x_3) T_1(x_2)
\end{displaymath}
\begin{displaymath}
+T_1(x_3) T_1(x_1) T_1(x_2) + T_1(x_3) T_1(x_2) T_1 (x_1)
\end{displaymath}
\begin{equation}
=\lambda^3 \Delta_3^-(x_1,x_2,x_3) :\Phi(x_1) \Phi(x_2) \Phi(x_3): + \mbox{\it{other graphs}},
\, \label{defdelta3minus}
\end{equation}
\begin{displaymath}
\Delta_{3,op}^{+}(x_1,x_2,x_3)=
+ T_1(x_1) T_2(x_2,x_3) + T_2(x_1, x_2) T_1(x_3) + T_1(x_2) T_2(x_1,x_3)
\end{displaymath}
\begin{displaymath}
-T_1(x_1) T_1(x_2) T_1(x_3) -  T_1(x_2) T_1 (x_1) T_1(x_3)
\end{displaymath}
\begin{equation}
=\lambda^3 \Delta_3^+(x_1,x_2,x_3) :\Phi(x_1) \Phi(x_2) \Phi(x_3): + \mbox{\it{other graphs}},
\, \label{defdelta3plus}
\end{equation}
and
\begin{displaymath}
\Delta_3^{op} (x_1,x_2,x_3) =
\Delta_{3,op}^{+}(x_1,x_2,x_3)+\Delta_{3,op}^{-}(x_1,x_2,x_3)
\end{displaymath}
\begin{displaymath}
=[T_1(x_1),T_2(x_2,x_3)]+[T_1(x_2), T_2(x_1,x_3)]
-[T_1(x_3), T_2(x_1,x_2)]
\end{displaymath}
\begin{displaymath}
+[T_1(x_3),T_1(x_1)T_1(x_2)] + [T_1(x_3), T_1(x_2)T_1(x_1)]
\end{displaymath}
\begin{equation}
=\lambda^3 \Delta_3(x_1,x_2,x_3) :\Phi(x_1) \Phi(x_2) \Phi(x_3): + \mbox{\it{other graphs}},
\, \label{defdelta3}
\end{equation}
where $\Delta_3^{op}$ and $\Delta_3$
have causal support as distributions in the sense that
\begin{equation}
\mbox{supp} \, \Delta_3^{(op)}(x_1,x_2,x_3) \subseteq \Gamma^+_3 \cup \Gamma^-_3,
\label{causalcond}
\end{equation}
where
\begin{equation}
 \Gamma^\pm_n=\{ (x_1,...,x_n) | x_n \in \mathds{R}^4, \,
(x_j-x_n) \in \overline{V}^\pm \, \forall j=1,...,n-1 \}.
\end{equation}
Therefore, only when {\emph{both}} arguments $x_1$ and $x_2$ are in the forward light cone or
{\emph{both}} arguments $x_1$ and $x_2$ are in the backward light cone
with respect to $x_3$, $(x_1,x_2,x_3)$ belongs to the support of
$\Delta_3^{(op)}$.
For the sake of brevity, only a short outline how one can prove the special
support property of $\Delta_3^{op}$ is given here by considering the
case where, e.g., $x_1 \in \overline{V}^+(x_3)$ and $x_2 \not\in \overline{V}^+(x_3)$.
In this case, one can assume that one is in a Lorentz system where
$x_1^0 > x_3^0 > x_2^0$, since the transformation properties of all $T_n$'s are fixed by
the representation of the Lorentz group on the Fock space of massless scalar particles.
Then one has, e.g., $T_2(x_2,x_3)=T(T_1(x_2) T_1(x_3))=T_1(x_3) T_1(x_2)$, and accordingly
for the full expression eq. (\ref{defdelta3})
\begin{displaymath}
\Delta_3^{op} (x_1,x_2,x_3) =
\end{displaymath}
\begin{displaymath}
+T_1(x_1) T_1(x_3) T_1(x_2) - T_1(x_3) T_1(x_2) T_1(x_1)
\end{displaymath}
\begin{displaymath}
+T_1(x_2) T_1(x_1) T_1(x_3) - T_1(x_1) T_1(x_3) T_1(x_2)
\end{displaymath}
\begin{displaymath}
-T_1(x_3) T_1(x_1) T_1(x_2) + T_1(x_1) T_1(x_2) T_1(x_3)
\end{displaymath}
\begin{displaymath}
+T_1(x_3) T_1(x_1) T_1(x_2) - T_1(x_1) T_1(x_2) T_1(x_3)
\end{displaymath}
\begin{equation}
+T_1(x_3) T_1(x_2) T_1(x_1) - T_1(x_2) T_1(x_1) T_1(x_3)=0. \label{delta3def}
\end{equation}
The same way one may check for all other cases that $\Delta_3^{op}$ vanishes whenever
the causal condition (\ref{causalcond}) is not fulfilled.
The general definition of causal distributions $\Delta_n^{op}(x_1,...,x_n)$
at higher orders can be found in \cite{GScharf,Aste}.

One should point out that
$x_3$ plays a distinguished role in the discussion above.
Whereas the time-ordered distribution $T_3(x_1,x_2,x_3)=T_3(x_{i_1},x_{i_2},x_{i_3})$
is invariant under a permutation $(1,2,3) \rightarrow (i_1,i_2,i_3)$ of the
arguments according to its definition in eqns. (\ref{stoer1},\ref{toprod},\ref{smoothedS}),
$\Delta_3^{op}(x_1,x_2,x_3)$ is not. However, the construction presented above
would go through with $x_1$ or $x_2$ as distinguished arguments.
The crucial point is that the causal support property of $\Delta_3^{op}$ and its C-number part
$\Delta_3$ will enable the calculation of a distribution which is retarded with
respect to $x_3$ via a double dispersion relation in momentum space, as will be shown
in the following. This retarded distribution ${\hat{\Delta}}_3^{ret}$
can be related directly to the vertex function in analogy to eq. (\ref{causal1}).
The vertex function again possesses the full permutation symmetry in configuration
space by construction.

The vertex function will be constructed below by performing the following calculational
steps. The Fourier transform $\hat{\Delta}_3(p,q)=\int d^4 x_1 d^4 x_2
\Delta_3(x_1,x_2,x_3) e^{ip(x_1-x_3)+iq(x_3-x_2)}$, which is independent of $x_3$
due to translation invariance (see also chapter 3.1 in \cite{GScharf}), is calculated
from the expression eq. (\ref{defdelta3}) in configuration space.
Then the corresponding retarded distribution $\Delta_3^{ret}(x_1,x_2,x_3) = \Theta(x_1-x_3)
\Theta(x_2-x_3) \Delta_3 (x_1,x_2,x_3)$ is evaluated in momentum space
by a dispersion relation analogously to eq. (\ref{disprel}),
and finally the vertex function is obtained directly from
analytic considerations. It is also straightforward to check that on the
operator level, the time-ordered product $T_3$ can be obtained from the retarded
(appropriately regularized) distribution $\Delta_3^{op,ret} (x_1,x_2,x_3)$
\begin{equation}
T_3(x_1,x_2,x_3)=\Delta_3^{op,ret} (x_1,x_2,x_3) -\Delta_{3,op}^{-}(x_1,x_2,x_3),
\label{vertexconfig}
\end{equation}
i.e. for the vertex distribution holds
\begin{equation}
V(x_1,x_2,x_3)=\Delta_3^{ret} (x_1, x_2, x_3) - \Delta_3^- (x_1,x_2,x_3) \label{3composition}
\end{equation}
in close analogy to eq. (\ref{causal1}).

\section{Calculation of $\hat{\Delta}_3$}
Starting from the first and second order tree diagram given by
\begin{displaymath}
T_2(x_1,x_2)=(-i)^2\frac{\lambda^2}{(3!)^2} T :\Phi^3(x_1): :\Phi^3(x_2):=
\end{displaymath}
\begin{equation}
-i\frac{\lambda^2}{4} :\Phi^2(x_1) \Phi^2(x_2): \Delta_F(x_1-x_2)
+ {\mbox{\emph{other graphs}}},
\end{equation}
one readily obtains for $\Delta_3$ in configuration space by exploiting
the relevant Wick contractions in eq. (\ref{defdelta3})
\begin{displaymath}
\Delta_3(x_1,x_2,x_3) = \Delta_3^{+}(x_1,x_2,x_3)+\Delta_3^{-}(x_1,x_2,x_3)=
\end{displaymath}
\vskip -0.8 cm
\begin{eqnarray}
+\Delta^+ (x_1-x_2) \Delta^+ (x_1-x_3) \Delta_F (x_2-x_3) \, &
-\Delta^+ (x_2-x_1) \Delta^+ (x_3-x_1) \Delta_F (x_2-x_3) & \nonumber \\
+\Delta^+ (x_2-x_1) \Delta^+ (x_2-x_3) \Delta_F (x_1-x_3) &
-\Delta^+ (x_1-x_2) \Delta^+ (x_3-x_2) \Delta_F (x_1-x_3) & \nonumber \\
+\Delta_F (x_1-x_2) \Delta^+ (x_1-x_3) \Delta^+ (x_2-x_3) &
-\Delta_F (x_1-x_2) \Delta^+ (x_3-x_1) \Delta^+ (x_3-x_2) & \nonumber \\
+\Delta^+ (x_3-x_1) \Delta^+ (x_3-x_2) \Delta^+ (x_1-x_2) &
-\Delta^+ (x_1-x_3) \Delta^+ (x_2-x_3) \Delta^+ (x_1-x_2) & \nonumber \\
+\Delta^+ (x_3-x_1) \Delta^+ (x_3-x_2) \Delta^+ (x_2-x_1) & 
-\Delta^+ (x_1-x_3) \Delta^+ (x_2-x_3) \Delta^+ (x_2-x_1) & \! \! \!. \label{delta3e}
\end{eqnarray}
Note that a Wick contraction {\em{without}} time-ordering leads to a $\Delta^+$-distribution
instead of a Feynman propagator.
Making use of $\Delta^+(-x)=-\Delta^-(x)$ and introducing the
advanced distribution $\Delta^{av}$ such that $\Delta(x)=\Delta^{ret}(x)-\Delta^{av}(x)$
and $\Delta^{ret}(-x)=-\Delta^{av}(x)$,  eq. (\ref{delta3e}) can be simplified
to
\begin{displaymath}
\Delta_3(x_1,x_2,x_3)=
\end{displaymath}
\begin{displaymath}
+\Delta^- (x_1-x_3) \Delta^+ (x_3-x_2) \Delta_F (x_1-x_2)
+\Delta^+ (x_1-x_3) \Delta^{ret} (x_3-x_2) \Delta^+ (x_1-x_2)
\end{displaymath}
\begin{equation}
+\Delta^- (x_1-x_3) \Delta^{av} (x_3-x_2) \Delta^+ (x_2-x_1) -
\{(x_1,x_2,x_3) \leftrightarrow (-x_2,-x_1,-x_3)\}.
\end{equation}
Performing the Fourier transform, one obtains
\begin{displaymath}
\hat{\Delta}_3(p,q)=\int d^4 x_1 d^4 x_2 \Delta(x_1,x_2,x_3)
e^{ip(x_1-x_3)+iq(x_3-x_2)}=(2 \pi)^{-4} \int d^4 k \times
\end{displaymath}
\begin{displaymath}
\bigl[ \hat{\Delta}^- (p-k) \hat{\Delta}^+ (q-k) \hat{\Delta}_F(k) +
\end{displaymath}
\begin{displaymath}
\hat{\Delta}^+ (p-k) \hat{\Delta}^{ret} (q-k) \hat{\Delta}^+(k)+
\end{displaymath}
\begin{equation}
\hat{\Delta}^- (p-k) \hat{\Delta}^{av} (q-k) \hat{\Delta}^+(-k) \bigr] -
\{ p \leftrightarrow q\}. \label{delta3momentum}
\end{equation}
From the integral eq. (\ref{delta3momentum}) one immediately derives the following
symmetry properties of $\hat{\Delta}_3$
\begin{displaymath}
\hat{\Delta}_3(p,q)=-\hat{\Delta}_3(q,p)=\hat{\Delta}_3(-p,-q),
\end{displaymath}
\begin{equation}
\hat{\Delta}_3(-p,q)=\hat{\Delta}_3(-q,p).
\end{equation}
As an illustration, the full calculation of the first Fourier integral
appearing in eq. (\ref{delta3momentum})
\begin{displaymath}
(2 \pi)^{-2} I_1(p,q)= (2 \pi)^{-4} \int d^4 k
\bigl[ \hat{\Delta}^- (p-k) \hat{\Delta}^+ (q-k) \hat{\Delta}_F(k) \bigr]
\end{displaymath}
\begin{equation}
=(2 \pi)^{-2} \int d^4 k\Theta(k^0-p^0) \delta((p-k)^2) \Theta(q^0-k^0) \delta((q-k)^2)
\frac{1}{k^2+i0}
\end{equation}
is presented here.
Introducing $P=p-q$ and using a new integration variable $k'=k-p$, one obtains
\begin{equation}
I_1(p,q)=\int \frac{d^4 k'}{(k'+p)^2+i0} \Theta (k'^0)\delta(k'^2) \Theta(-P^0-k'^0)
\delta((P+k')^2). \label{int1}
\end{equation}
The first $\Theta$ and $\delta$ distributions can be written as $\delta(k^0-|\vec{k}|)/2|\vec{k}|$
($k'$ is replaced by $k$ in the following).
Taking into account the second $\Theta$ distribution one easily derives $P^0 \le-k^0 \le 0$,
otherwise integral above would vanish.
For spacelike $P$, $P^2 < 0$, there exists a Lorentz frame with $P^0 > 0$
such that $I_1$ also vanishes.
For timelike $P$, there exists a Lorentz frame where $\vec{P}=\vec{0}
=\vec{p}-\vec{q}$. The last $\delta$ distribution in eq. (\ref{int1}) then
implies $(P^0+k^0)^2-\vec{k}^2=0$, and combined with $k^0=|\vec k|$ from the $\delta$
distribution $\delta(k^0-|\vec k|)$ one can replace $k^0$ by $-P^0/2$ in
the Feynman propagator appearing in eq. (\ref{int1}) after performing the trivial integral over
$k^0$.
For timelike $P$, there remains to calculate with
$\delta((P^0+k^0)^2-\vec{k}^2)$ replaced by $\delta(P_0^2+2P_0 | {\vec{k}}|)=
\delta(|\vec{k}|-\frac{|P^0|}{2})/2 | P^0|$
\begin{equation}
I_1(p,q)=\Theta(-P^0) \int \frac{d^3k}{2|\vec{k}|} \frac{\delta(|\vec{k}|-\frac{|P^0|}{2})}
{2|P^0| [(k+p)^2+i0]},
\end{equation}
with $k$ in the denominator above given by $k=(-P^0/2,{\vec{k}})$, such that
one can write
\begin{equation}
(k+p)^2=(k^0+p^0)^2-(\vec{k}+\vec{p})^{\, 2}=(-\frac{1}{2} P^0 + p^0)^2
 - \vec{k}^2 -\vec{p}^{\, 2} - 2 |\vec{k}| |\vec{p} \, | \cos \vartheta,
\end{equation}
and one has $|\vec{k}|=|P^0|/2$. Hence ($P^0=P_0$),
\begin{displaymath}
I_1(p,q)=\frac{2 \pi}{8} \Theta(-P^0)\Theta(P_0^2) \int \limits_{-1}^{+1} \frac{d \cos \vartheta}
{(-P^0/2+p^0)^2 - \vec{k}^2 -\vec{p}^{\, 2} -2 |\vec{k}||\vec{p} \, | \cos \vartheta + i0}
\end{displaymath}
\begin{displaymath}
=-\frac{\pi}{4} \frac{\Theta(-P^0)\Theta(P_0^2)}{|P^0||\vec{p} \, |}
\log \Biggl( \frac{(-P^0/2+p^0)^2-(|\vec{k}|+|\vec{p} \, |)^2 + i0}
{(-P^0/2+p^0)^2-(|\vec{k}|-|\vec{p} \, |)^2 + i0} \Biggr)
\end{displaymath}
\begin{equation}
=-\frac{\pi}{4} \frac{\Theta(-P^0)\Theta(P_0^2)}{|P^0||\vec{p} \, |}
\log \Biggl( \frac{p^2-P^0 p^0 - |\vec{p} \, | |P^0| +i0}
{p^2-P^0 p^0 + |\vec{p} \, | |P^0| +i0} \Biggr) .
\label{igen}
\end{equation}
The distribution $\Theta(P_0^2)$ inserted in eq. (\ref{igen}) does not
affect the result in the rest frame of $P$, but will account below for the fact that
$I_1(p,q)$ vanishes for spacelike $P$ when written in covariant form $\Theta(P^2)$.
Investigating the behavior of the expression $|\vec{p} \, ||P^0|$ under Lorentz transformations
one finds that the covariant form of $|\vec{p} \, ||P^0|$ is given by $\sqrt{N}$, where $N$ is
the triangle function
\begin{equation}
N \equiv (Pp)^2-P^2p^2=(pq)^2-p^2q^2,
\end{equation}
such that one arrives at
\begin{displaymath}
I_1(p,q)=-\frac{\pi}{4} \frac{\Theta(-P^0)\Theta(P^2)}{\sqrt{N}}
\log \Biggl( \frac{pq-\sqrt{N}+i0}{pq+\sqrt{N}+i0} \Biggr)
\end{displaymath}
\begin{equation}
=\frac{\pi}{4} \frac{\Theta(-P^0)\Theta(P^2)}{\sqrt{N}}
\log \Biggl( \frac{pq+\sqrt{N}}{pq-\sqrt{N}} -i0 \Biggr).
\end{equation}
It is left to the reader as a simple exercise to show that lightlike $P$ do not contribute
to $I_1(p,q)$ in a distributional sense.
Calculating the remaining two integrals along the same lines as above and subtracting
the resulting expressions with $p$ and $q$ interchanged, one obtains
\begin{displaymath}
\hat{\Delta}_3(p,q)= \frac{\pi}{4 (2 \pi)^2}
\Biggl\{ -\frac{\mbox{sgn}(P^0) \Theta (P^2)}{\sqrt{N}} \log \Biggl|
\frac{pq+\sqrt{N}}{pq-\sqrt{N}} \Biggr|
\end{displaymath}
\begin{equation}
+\frac{\mbox{sgn}(q^0) \Theta (q^2)}{\sqrt{N}} \log \Biggl|
\frac{p^2-pq+\sqrt{N}}{p^2-pq-\sqrt{N}} \Biggr|-
\frac{\mbox{sgn}(p^0) \Theta (p^2)}{\sqrt{N}} \log \Biggl|
\frac{q^2-pq+\sqrt{N}}{q^2-pq-\sqrt{N}} \Biggr| \Biggl\}. \label{delta3result}
\end{equation}
Note that all imaginary parts of the integrals appearing in eq. (\ref{delta3momentum})
mutually cancel, since $\hat{\Delta}_3$ is real due to its antisymmetry properties
in configuration space described above.

\section{Calculation of $\hat{\Delta}_3^{ret}$}
In close analogy to eq. (\ref{disprel}), $\hat{\Delta}_3^{ret}(p,q)$ is calculated in the
following for 
$p \in V^+$ and $q \in V^-$. The result for arbitrary $p$, $q$ can be obtained by analytic
continuation, since $\hat{\Delta}_3^{ret}(p,q)$ is the boundary value of an analytic function,
regular in $p \in \mathds{R}^{4}+iV^+$, $q \in \mathds{R}^{4}+iV^-$ (see also \cite{Reeds}).
The procedure is in two steps. First, the integral
\begin{equation}
\hat{\Delta}_{3,p}(p,q)=\frac{i}{2 \pi} \int \limits_{-\infty}^{+\infty}
dt \frac{\hat{\Delta}_3(tp,q)}{1-t+i0} \label{split1}
\end{equation}
is evaluated for $p \in V^+$. This corresponds to the calculation of
$\Theta(x_1^0-x_3^0) \Delta_3(x_1,x_2,x_3)$ in con\-fi\-gu\-ra\-ti\-on space in
the Lorentz system where $\vec{p}=(p^0,\vec{0})$.
Then $\hat{\Delta}_3^{ret}(p,q)$ is obtained by calculating
\begin{equation}
\hat{\Delta}_3^{ret}(p,q)=\frac{i}{2 \pi} \int \limits_{-\infty}^{+\infty} dt
\frac{\hat{\Delta}_{3,p} (p,tq)}{1-t+i0}=
-\frac{1}{(2 \pi)^2} \int \limits_{-\infty}^{+\infty} dt \int \limits_{-\infty}^{+\infty} ds
\frac{\hat{\Delta}_3(tp,sq)}{(1-t+i0)(1-s+i0)}, \label{doublesplit}
\label{split2}
\end{equation}
corresponding to the multiplication of $\Theta(x_1^0-x_3^0) \Delta_3(x_1,x_2,x_3)$ with
$\Theta(x_2^0-x_3^0)$ in configuration space.
Note that the well-known splitting formula \cite{GScharf}
\begin{equation}
\Delta_3^{ret}(p,q)=\frac{i}{2 \pi} \int \limits_{-\infty}^{+\infty} dt
\frac{\hat{\Delta}_3 (tp,tq)}{1-t+i0}, \label{naivesplitting}
\end{equation}
is infrared divergent in the case of a totally massless vertex, while the double dispersion
relation eq. (\ref{doublesplit}) is not.

For the actual calculations one may start by considering the first logarithm in
eq. (\ref{delta3result})
\begin{equation}
I_{1,p}(p,q)=-\frac{i \pi}{4 (2 \pi)^3} \int \limits_{-\infty}^{+\infty} dt
\frac{\mbox{sgn}(tp^0-q^0) \Theta((tp-q)^2)}{(1-t+i0)|t|\sqrt{N}}
\log \Biggl| \frac{tpq+|t|\sqrt{N}}{tpq-|t|\sqrt{N}} \Biggr|. \label{I1p}
\end{equation}
The real part of the integral (\ref{I1p}) can be written down immediately ($P=p-q$)
\begin{equation}
Re[I_{1,p}(p,q)]=-\frac{\pi^2}{4(2 \pi)^3} \frac{\mbox{sgn}(P^0) \Theta(P^2)}{\sqrt{N}}
\log \Biggl| \frac{pq+\sqrt{N}}{pq-\sqrt{N}} \Biggr|.
\end{equation}
The Lorentz system shall be chosen such that
$p \rightarrow (p^0,\vec{0})$. Since
$p \in V^+$ (and $q$ arbitrary), one has to evaluate the Cauchy principal value
\begin{equation}
Im[I_{1,p}(p,q)]=-\frac{i \pi}{4 (2 \pi)^3} \log \Biggl| \frac{pq+\sqrt{N}}{pq-\sqrt{N}} \Biggr|
 \int \limits_{-\infty}^{+\infty} dt
\frac{\mbox{sgn}(tp^0-q^0) \Theta(((tp^0,\vec{0})-q)^2)}{t(1-t)}. \label{analogo}
\end{equation}
From the $\Theta$ distribution, one obtains integral limits via
\begin{equation}
t^2 p_0^2 - 2t p^0 q^0 + q^2=0 \Rightarrow t_{1,2}=\frac{1}{p^0}
\Bigl( q^0 \pm \sqrt{q_0^2-q^2} \Bigr)=\frac{1}{p^0}
(q^0 \pm |\vec{q} \, |),
\end{equation}
and taking into account that $\mbox{sgn}(t_{1,2}p^0-q^0)=\mbox{sgn}(\pm|\vec{q} \, |)$,
one obtains from
\begin{displaymath}
-\int \limits_{-\infty}^{t_2} \frac{dt}{t(1-t)}+
\int \limits_{t_1}^{\infty} \frac{dt}{t(1-t)}=\log \Biggl| \frac{(1-t_1)(1-t_2)}{t_1 t_2} \Biggr|
\end{displaymath}
\begin{equation}
=\log \Biggl| \frac{((p^0,\vec{0})-q^2}{q^2} \Biggr| =
\log \Biggl| \frac{(p-q)^2}{q^2} \Biggr|
\end{equation}
the first 'splitting' result
\begin{equation}
I_{1,p}(p,q)=\frac{i \pi}{4 (2 \pi)^3 \sqrt{N}}
\log \Biggr| \frac{pq+\sqrt{N}}{pq-\sqrt{N}} \Biggr| \log \Biggl| \frac{q^2}{(p-q)^2} \Biggr|
-\frac{\pi^2}{4(2 \pi)^3} \frac{\mbox{sgn}(P^0) \Theta(P^2)}{\sqrt{N}}
\log \Biggl| \frac{pq+\sqrt{N}}{pq-\sqrt{N}} \Biggr|.
\end{equation}
As further step one may calculate
\begin{equation}
I_{1,p,q}=\frac{i}{2 \pi} \int \limits_{-\infty}^{+\infty} dt \frac{I_{1,p}(p,tq)}{1-t+i0}.
\end{equation}
The terms generated by the $i0$-term can be written down immediately
\begin{equation}
\frac{i \pi^2}{4 (2\pi)^4 \sqrt{N}} \log \Biggl| \frac{pq+\sqrt{N}}{pq-\sqrt{N}} \Biggr|
\log \Biggl| \frac{q^2}{(p-q)^2} \Biggr| - \frac{\pi^3}{4 (2 \pi)^4 \sqrt{N}} \log
\Biggl| \frac{pq+\sqrt{N}}{pq-\sqrt{N}} \Biggr|. \label{izeros}
\end{equation}
The remaining non-trivial integrals are
\begin{equation}
i_1=-\frac{\pi}{4 (2 \pi)^4 \sqrt{N}} \log 
\Biggr| \frac{pq+\sqrt{N}}{pq-\sqrt{N}} \Biggr| \int \limits_{-\infty}^{+\infty}
\frac{dt}{t(1-t)} \log \Biggl| \frac{t^2 q^2}{(p-tq)^2} \Biggr| \label{i_1}
\end{equation}
and
\begin{equation}
i_2=-\frac{i \pi^2}{4 (2 \pi)^4 \sqrt{N}} \log \Biggl| \frac{pq+\sqrt{N}}{pq-\sqrt{N}} \Biggr|
\int \limits_{-\infty}^{+\infty}
\frac{dt}{t(1-t)} \mbox{sgn}(tP^0) \Theta(t^2 P^2). \label{i_2}
\end{equation}
Integral $i_1$ can easily be solved by the help of the identity
\begin{equation}
\lim_{R \rightarrow \infty} \int \limits_{-R}^{+R} dt \frac{\log|t-\alpha|}{t-\beta}
=\frac{\pi^2}{2} \mbox{sgn}(\beta-\alpha),
\end{equation}
and one obtains
\begin{equation}
i_1=\frac{\pi^3}{4 (2 \pi)^4 \sqrt{N}} \log 
\Biggr| \frac{pq+\sqrt{N}}{pq-\sqrt{N}} \Biggr|,
\end{equation}
i.e. $i_1$ cancels the second term in eq.(\ref{izeros}). Integral $i_2$ has a
structure completely analogous to eq. (\ref{analogo}) and is given by
\begin{equation}
\frac{i \pi^2}{4 (2 \pi)^3 \sqrt{N}}
\log \Biggr| \frac{pq+\sqrt{N}}{pq-\sqrt{N}} \Biggr| \log \Biggl| \frac{p^2}{(p-q)^2} \Biggr|
\end{equation}
such the compact result ($p \in V^+$, $q \in V^-$)
\begin{equation}
I_{1,p,q}=\frac{i \pi^2}{4 (2 \pi)^4 \sqrt{N}}
\log \Biggl( \frac{pq+\sqrt{N}}{pq-\sqrt{N}} \Biggr) 
\log \Biggl( \frac{p^2q^2}{(p-q)^2} \Biggr)
\end{equation}
is obtained.

The integrals which appear in the calculation of the remaining parts
of $\hat{\Delta}_3^{ret} (p,q)$, namely $I_{2,p,q}$ and $I_{3,p,q}$,
can all be expressed by the help of the Spence function (or dilogarithm) $Li(z)$,
which is defined by
\begin{equation}
Li(z)=\int \limits_{0}^{z} \frac{dt}{t} \log(1-t), \label{spence}
\end{equation}
where the integral contour in eq. (\ref{spence}) must be equivalent to a straight line
from $0$ to $z$ avoiding $[1,+\infty) \subset \mathds{R}$.
This is due to the fact that all relevant integrals
can be reduced directly to the form $(a,b,c,d \in \mathds{R})$
\begin{equation}
\int \limits^{x} dt \frac{\log| at+b|}{ct+d}=-\frac{1}{c} L \Biggl(
\frac{a(cx+d)}{ad-bc} \Biggr) + \frac{1}{c}
\log \Biggl| \frac{ad-bc}{c} \Biggr| \log \Biggl| \frac{a(cx+d)}{ad-bc} \Biggr|,
\end{equation}
where $L(x)$ is the real Spence function, defined for real arguments
$x \! \in \! \mathds{R}$ by
\begin{equation}
L(x) = \frac{1}{2} (Li(x+i0)+Li(x-i0)).
\end{equation}
For the sake of brevity, the results are presented here
\begin{displaymath}
I_{2,p}(p,q)=\frac{i \pi}{4 (2 \pi)^3} \int \limits_{-\infty}^{+\infty} dt
\frac{\mbox{sgn}(q^0) \Theta(q^2)}{(1-t+i0)|t|\sqrt{N}}
\log \Biggl| \frac{t^2p^2-tpq+|t|\sqrt{N}}{t^2p^2-tpq-|t|\sqrt{N}} \Biggr|
\end{displaymath}
\begin{equation}
=\frac{\pi^2}{4(2 \pi)^3} \frac{\mbox{sgn}(q^0) \Theta(q^2)}{\sqrt{N}}
\log \Biggr| \frac{p^2-pq+\sqrt{N}}{p^2-pq-\sqrt{N}} \Biggr|
-\frac{i \pi^3}{4 (2 \pi)^3} \frac{\Theta(q^0) \Theta(q^2) \Theta(-P^2)}{\sqrt{N}},
\end{equation}
\begin{displaymath}
I_{2,p,q}(p,q)=-\frac{i \pi^2}{4(2 \pi)^4 \sqrt{N}} \Biggl[ 2 Li \Biggl( \frac{p^2-pq-\sqrt{N}}
{-pq-\sqrt{N}} \Biggr) - 2Li \Biggl( \frac{p^2-pq+\sqrt{N}}{-pq+\sqrt{N}} \Biggr)
\end{displaymath}
\begin{displaymath}
+ \log^2 \Biggl( \frac{-pq+\sqrt{N}}{p^2} \Biggr) - \log^2 \Biggl( \frac{-pq-\sqrt{N}}{p^2}
\Biggr) \Biggr]
\end{displaymath}
\begin{equation}
-\frac{\pi^3}{4(2 \pi)^4 \sqrt{N}} \Biggl[ \log \Biggl( \frac{pq-p^2-\sqrt{N}}{pq-p^2+\sqrt{N}}
\Biggr) + \log \Biggl( \frac{pq-q^2-\sqrt{N}}{pq-q^2+\sqrt{N}} \Biggr)
+ \log \Biggl( \frac{pq+\sqrt{N}}{pq-\sqrt{N}} \Biggr) \Biggr], \label{I2split}
\end{equation}
and $I_{3,p,q}$ ist given by $I_{3,p,q}(p,q)=I_{2,p,q}(q,p)$.
Note that the last three logarithmic terms in eq. (\ref{I2split}) are superfluous. They
indeed appear in the actual calculations, however, they combine to zero, since
\begin{equation}
\Biggl( \frac{pq-p^2-\sqrt{N}}{pq-p^2+\sqrt{N}}
\Biggr) \Biggl( \frac{pq-q^2-\sqrt{N}}{pq-q^2+\sqrt{N}} \Biggr)
\Biggl( \frac{pq+\sqrt{N}}{pq-\sqrt{N}} \Biggr) = 1.
\end{equation}
It must also be pointed out that the Spence function obeys very many identities. E.g.,
the real Spence function satisfies
\begin{displaymath}
L(1-x)=-L(x)+\log|x| \log|1-x|-\frac{\pi^2}{6}
\end{displaymath}
\begin{displaymath}
=L(1/x)-\frac{1}{2} \log^2 |x| + \log |x| \log |1-x| +
\left\{ \begin{array}{rcc} -\pi^2/3 & : & x<0  \\ +\pi^2/6 & : &  x>0
\end{array} \right.
\end{displaymath}
\begin{equation}
=-L(1-1/x)+\frac{1}{2} \log^2 |x| +\left\{ \begin{array}{rcc} -\pi^2/2 & : & x<0  \\
0 & : &  x>0
\end{array} \right. \quad . \label{identities}
\end{equation}
Looking for a compact expression for $\hat{\Delta}_3^{ret}$, one finds
\begin{displaymath}
\hat{\Delta}_3^{ret}(p,q)=\sum \limits_{i=1}^{3} I_{i,p,q} (p,q)
\end{displaymath}
\begin{displaymath}
=\frac{i \pi^2}{2(2 \pi)^4 \sqrt{N}} \Biggl[ 2 Li \Biggl( \frac{p^2-pq-\sqrt{N}}{p^2}
\Biggr) - 2 Li \Biggl( \frac{p^2-pq+\sqrt{N}}{p^2} \Biggr)
\end{displaymath}
\begin{equation}
+\log \Biggl( \frac{pq+\sqrt{N}}{pq-\sqrt{N}} \Biggr)
\log \Biggl( \frac{p^2}{(p-q)^2} \Biggr) \Biggr]. \label{retarded}
\end{equation}
This expression can be transformed with the help of eq. (\ref{identities}) and
the observation that the arguments of the Spence function can be written in many different ways
\begin{displaymath}
\frac{p^2-pq-\sqrt{N}}{-pq-\sqrt{N}}=1-\frac{pq-\sqrt{N}}{q^2}=1-\frac{p^2}{pq+\sqrt{N}},
\end{displaymath}
\begin{equation}
\frac{p^2-pq+\sqrt{N}}{-pq+\sqrt{N}}=1-\frac{pq+\sqrt{N}}{q^2}=1-\frac{p^2}{pq-\sqrt{N}}.
\end{equation}
Note that $\hat{\Delta}_3^{ret}(p,q)$ is purely imaginary for $p \in V^+, q \in V^-$.
This insight can also be obtained directly from the naive
splitting formula eq. (\ref{naivesplitting}), where only the imaginary part is infrared
divergent. There, one obtains for the real part of $\hat{\Delta}_3^{ret}(p,q)$ (see also eq.
(\ref{I2split}))
\begin{equation}
-\frac{2 \pi^3}{4(2 \pi)^4 \sqrt{N}} \Biggl[ \log \Biggl( \frac{pq-p^2-\sqrt{N}}{pq-p^2+\sqrt{N}}
\Biggr) + \log \Biggl( \frac{pq-q^2-\sqrt{N}}{pq-q^2+\sqrt{N}} \Biggr)
+ \log \Biggl( \frac{pq+\sqrt{N}}{pq-\sqrt{N}} \Biggr) \Biggr]=0.
\end{equation}
The retarded distribution $\hat{\Delta}_3^{ret}(p,q)$ can be obtained for
{\emph{arbitrary}} $p$, $q$ from eq. (\ref{retarded}) by analytic continuation,
since $\hat{\Delta}_3^{ret}(p,q)$ is the boundary
value of the corresponding function analytic for $p \in \mathds{R}^4+ i V^+$ and
$q \in \mathds{R}^4+iV^-$.
Furthermore, the full vertex function is finally obtained from eqns. (\ref{vertexconfig},
\ref{3composition})
\begin{equation}
\hat{V}(p,q)=\hat{\Delta}_3^{ret}(p,q)-\hat{\Delta}_3^- (p,q),
\end{equation}
with
\begin{displaymath}
\hat{\Delta}_3^{-} (p,q)= \frac{\pi}{4 (2 \pi)^2} \Biggl\{
\Biggl( \frac{\Theta (-P^0) \Theta (P^2)}{\sqrt{N}} \log \Biggl(
\frac{pq+\sqrt{N}}{pq-\sqrt{N}} -i0 \Biggr)
\end{displaymath}
\begin{equation}
+\frac{\Theta (q^0) \Theta (q^2)}{\sqrt{N}} \log \Biggl(
\frac{p^2-pq+\sqrt{N}}{p^2-pq-\sqrt{N}} -i0 p^2 \Biggr)+
\frac{\Theta (-p^0) \Theta (p^2)}{\sqrt{N}} \log \Biggl(
\frac{q^2-pq+\sqrt{N}}{q^2-pq-\sqrt{N}} -i0 q^2 \Biggr) \Biggl\}.
\end{equation}
The result for $\hat{\Delta}_3^{-} (p,q)$ is given here with the remark that
$\hat{\Delta}_3^{-} (p,q)$ and $\hat{\Delta}_3(p,q)$ can indeed be calculated in parallel.
A detailed calculation of
$\hat{\Delta}_3^{\pm}$ and $\hat{\Delta}_3$ can also be found in \cite{GScharf},
chapter 3.8, where the vertex function appearing in QED with massive electrons
is treated. The result for the massless case can be easily obtained by
setting the electron mass equal to zero, since no infrared problems are present in
$\hat{\Delta}_3^{\pm}$ and $\hat{\Delta}_3$.

For $p \in V^+$ and $q \in V^-$, $\hat{\Delta}_3^{-}(p,q)$ vanishes and one can finally
write down in a 'sloppy' style the result that can be found in the literature
\cite{Tarasov, Ball, Kizilersu}
\begin{displaymath}
\int \frac{d^4 k}{[(p-k)^2][(q-k)^2][k^2]}
\end{displaymath}
\begin{equation}
=\frac{i \pi^2}{2 \sqrt{N}} \Biggl[ 2 Li \Biggl( \frac{p^2-pq-\sqrt{N}}{p^2}
\Biggr) - 2 Li \Biggl( \frac{p^2-pq+\sqrt{N}}{p^2} \Biggr)+
\log \Biggl( \frac{pq+\sqrt{N}}{pq-\sqrt{N}} \Biggr)
\log \Biggl( \frac{p^2}{(p-q)^2} \Biggr) \Biggr]. \label{vresult}
\end{equation}
Note that in the result given in \cite{Ball}, a factor of two is missing.
Eq. (\ref{vresult}) is also equivalent to the forms derived in \cite{Davy2,Davy3}.

\begin{acknowledgements}
This work was supported by the Swiss National Science Foundation.
\end{acknowledgements}

\end{document}